\documentclass[pre,onecolumn,showpacs,superscriptaddress]{revtex4-1}
\usepackage[T1]{fontenc}
\usepackage{graphicx}
\usepackage{geometry} 
\usepackage{color,amssymb}

\begin{document}

\title{Reconstruction of multiplex networks with correlated layers}

\author{Valerio Gemmetto}
\email{gemmetto@lorentz.leidenuniv.nl}
\affiliation{Instituut-Lorentz for Theoretical Physics, Leiden Institute of Physics, University of Leiden, Niels Bohrweg 2, 2333 CA Leiden, The Netherlands}

\author{Diego Garlaschelli}
\email{garlaschelli@lorentz.leidenuniv.nl}
\affiliation{Instituut-Lorentz for Theoretical Physics, Leiden Institute of Physics, University of Leiden, Niels Bohrweg 2, 2333 CA Leiden, The Netherlands}

\begin{abstract}
The characterization of various properties of real-world systems requires the knowledge of the underlying network of connections among the system's components. Unfortunately, in many situations the complete topology of this network is empirically inaccessible, and one has to resort to probabilistic techniques to infer it from limited information.
While network reconstruction methods have reached some degree of maturity in the case of single-layer networks (where nodes can be connected only by one type of links), the problem is practically unexplored in the case of multiplex networks, where several interdependent layers, each with a different type of links, coexist.
Even the most advanced network reconstruction techniques, if applied to each layer separately, fail in replicating the observed inter-layer dependencies making up the whole coupled multiplex.
Here we develop a methodology to reconstruct a class of correlated multiplexes which includes the World Trade Multiplex as a specific example we study in detail.
Our method starts from any reconstruction model that successfully reproduces some desired marginal properties, including node strengths and/or node degrees, of each layer separately. It then introduces the minimal dependency structure required to replicate an additional set of higher-order properties that quantify the portion of each node's degree and each node's strength that is shared and/or reciprocated across pairs of layers.
These properties are found to provide empirically robust measures of inter-layer coupling.
Our method allows joint multi-layer connection probabilities to be reliably reconstructed from marginal ones, effectively bridging the gap between single-layer properties and truly multiplex information.
\end{abstract}

\maketitle

\section{Introduction}

In the last twenty years, the study of complex networks acquired importance as it could significantly increase our understanding of many real-world systems~\cite{barabasi1999,albert2002,boccaletti2006}, ranging from the global airport infrastructure~\cite{guimera2004} to biological systems like the brain~\cite{bullmore2009}. Indeed, it is easy to realize that several systems, including the ones just mentioned, share a common abstract representation in terms of nodes connected by links, i.e. in terms of graphs or networks.

However, a more careful analysis shows that a simple network representation is often not enough to fully capture the whole complexity of the aforementioned systems~\cite{buldyrev2010}. For instance, the presence of different airline companies significantly affects the air transportation landscape~\cite{cardillo2013,cardillo2012}. Similarly, the human body can be thought of as a set of interdependent networks where several complex physiological systems, e.g. the nervous and the cardiovascular ones, constantly interact~\cite{bashan2012}.

For this reason, the concepts of multiplex and interdependent networks have been developed. In a multiplex network, a given set of nodes is connected through different modes of interactions; the system is therefore represented as a coloured-edge or layered graph~\cite{kivela2014}, where each layer contains the same set of ``replica nodes''.
Interdependent networks are instead composed of two or more interconnected networks, where each node of any graph is dependent on one or more nodes belonging to the other(s)~\cite{buldyrev2010}.

Several studies have focused on the analysis of structural aspects of these multi-graphs~\cite{battiston2014,nicosia2015,dedomenico2013}.
In particular, the analysis of the overlap between layers of a multiplex network can provide valuable information in order to better understand some dynamical processes that occur on top of those systems~\cite{gomez2013,cozzo2013} or possible failure cascades~\cite{buldyrev2010}. 
Moreover, the presence of dipendencies between layers crucially affects the systemic risk associated to these networks, for instance in the case of financial or economic systems~\cite{bargigli2014,poledna2015}.
It must be pointed out that, in order to study the aforementioned dynamical processes, the full graph structure is required, even in the case of monoplex networks. In general, however, confidentiality issues or limitedness of the topological information may not allow the knowledge of the entire network, but only that of partial information about the nodes (for instance, the degrees of all or some of the vertices, or the strengths and the density). Various \emph{network reconstruction} methods have therefore been developed, in order to successfully infer the full topological structure of graphs starting from incomplete information~\cite{clauset2008,mastromatteo2012,musmeci2013,caldarelli2013,mastrandrea2014bis,cimini1,cimini2,cimini3}.
Unfortunately, the current methodologies are applicable only to single-layer networks, leaving an important gap open in the study of multiplex networks. 
If these techniques were applied to each layer of a multiplex separately, they would by construction fail in replicating the empirical coupling between layers. 

Our main goal in the present paper is that of developing a satistifactory methodology for the reconstruction of multiplex networks from partial information.
Our approach is guided by the following consideration.
Clearly, a single-layer network can be seen as a particularly simple case of a multiplex, i.e. in the limit when the number of layers is one.
Then, from an entirely general point of view, a method to reconstruct multiplex networks may fail as a result of (a combination of) two factors. On one hand, the method may be unsuccessful because the properties of (some of) the layers are incorrectly reconstructed. This may be due to the method failing on each layer separately, a circumstance that strongly indicates an intrinsic unreliability of the reconstruction model itself, even when applied in the single-layer limit.
On the other hand, the method may succeed in replicating the marginal properties of each layer separately, while it may fail in replicating the interdependencies among layers.  
In the former case we do not learn anything useful about whether and how the method can be improved. By contrast, the latter situation is quite informative, as it indicates that, if the reconstruction model could be generalized in such a way that its marginal single-layer properties are maintained, while at the same time its inter-layer ones are made more realistic, then it would become an acceptable method for reconstructing multiplexes with coupled layers.

Following the above reasoning, we put ourselves in the latter situation and assume that the empirical multiplex is taken from a class of multiplex networks for which a `marginal' method capable of reliably reconstructing each layer separately exists.
Then, we investigate how a generalized and coupled multiplex method with the same marginal properties can be constructed.
Building on the recent literature on single-layer network reconstruction methods, we select the World Trade Multiplex (WTM) as the ideal empirical candidate for our analysis.
The nodes of this multiplex are countries of the world, whereas links represent trade relationships, disaggregated into different commodities.
Each commodity gives rise to a separate layer.
The links in each layer are in principle directed (from the exporter to the importer) and weighted (by the dollar value of the trade relationship), even though they are often projected into undirected and/or unweighted ones.  
The empirical properties of the WTM have been studied extensively~\cite{barigozzi2010,squartini2011I,squartini2011II,mastrandrea2014,gemmetto2015,gemmetto2016}.
If all the commodities are aggregated together, one obtains a single-layer projection documenting the total trade fluxes among countries~\cite{serrano2003,garlaschelli2005,squartini2011I,squartini2011II}.
In the representation considered here, we use data from Ref.~\cite{Comtrade,Baci} reporting $N = 207$ countries trading in $M = 96$ different commodities, each representing a given layer of the multiplex.

The WTM fulfills our criterion stated above, because it has been shown that each of its layers is very closely replicated by a model that takes only local node information as input. 
Indeed, the purely binary structure of each layer of the WTM can be replicated starting from the knowledge of the degree of each node in that layer~\cite{squartini2011I} (Binary Configuration Model~\cite{squartini2011,squartini2015}), while the weighted structure can be successfully replicated from the knowledge of both the strength and the degree of each node in that layer~\cite{mastrandrea2014} (Enhanced Configuration Model~\cite{mastrandrea2014bis,squartini2015}).
More relaxed reconstruction models~\cite{cimini1,cimini2,cimini3}, which are discussed later in the paper, have also been shown to successfully replicate the properties of the World Trade network.
At the same time, it has been shown that the knowledge of the strength and degree of each node in each layer is not enough to replicate the coupling between layers~\cite{gemmetto2015,gemmetto2016}, illustrating that even if the marginal reconstruction method is successful in each and every layer separately, it fails in replicating the multiplex as a whole.

Our strategy in this paper is that of devising a way to preserve the good marginal properties of single-layer reconstruction methods, while at the same time introducing a minimal but effective coupling such that, additionally, various robust inter-layer properties of the multiplex are also replicated. 
The structure of the paper is as follows. In sec.~\ref{sec:preliminaries} we introduce some preliminary concepts that constrain the range of possible multiplex reconstruction models. In sec.~\ref{sec:binary} we focus on the case of binary multiplexes (both undirected and directed) and develop a multiplex reconstruction method in that case.
In sec.~\ref{sec:weighted} we move on to weighted multiplexes (again, both undirected and directed) and develop the weighted counterpart of the reconstruction method.
Finally, in sec.~\ref{sec:conclusions} we make some concluding remarks.



\section{Preliminaries\label{sec:preliminaries}}
This section establishes some useful criteria which constrain the features of the multiplex reconstruction model we are after.
 
\subsection{Beyond inter-layer degree correlations}
To reliably reconstruct a multiplex, we need to identify useful target properties that accurately capture the inter-layer coupling.
Various notions of inter-layer overlap have  been developed in the literature, for instance in terms of \emph{correlation of layer activity}~\cite{nicosia2015} and \emph{overlapping degree}~\cite{battiston2014}.
In single-layer networks, degree correlation is usually computed by looking at the average degree of the first neighbours of a node having a certain degree (\emph{average nearest neighbour degree}). In the same spirit, notions of multiplex assortativity or inter-layer degree correlation have been developed~\cite{lee2012,nicosia2015,nicosia2013}. The inter-layer degree correlation function has been defined as:
\begin{equation}
\overline{k}^\alpha \left( k^\beta \right) = \sum_{k^\alpha} k^\alpha P \left( k^\alpha | k^\beta \right)
\label{assort_old}
\end{equation}
where $ P \left( k^\alpha | k^\beta \right) $ is the probability that a node having a given degree $ k^\beta $ in layer $ \beta $ has degree $ k^\alpha $ in layer $ \alpha $. 

We have recently shown that the above quantity is unfortunately not informative about the component of inter-layer coupling that is not due to the degree distribution of the various layers~\cite{gemmetto2015}.
For instance, if the same node is a hub in multiple layers (a property that gives rise to positive inter-layer assortativity), it will automatically produce a significant overlap of links across these layers, even if links in different layers are drawn completely independently. Such an overlap should therefore not be taken as a genuine measure of statistical dependency across layers.
This spurious effect increases with increasing intra-layer density and increasing heterogeneity of local node properties like degrees and strengths.
In order to detect `true' inter-layer dependencies that are not merely explained by chance, density, or by the local properties of individual nodes, one can construct maximum-entropy null models of multiplexes with independent layers and given node properties~\cite{garlaschelli2008,squartini2011,bianconi2013}. 
In these null models, in each layer every node has - on average - the same degree (for binary networks), or strength (for weighted networks), that it has in the real multiplex~\cite{gemmetto2015}. 
Apart from these constraints, the maximum-entropy multiplex ensemble is completely random and no dependency is introduced among layers.
The expectation values of the multiplexity over the null ensemble can be calculated exactly and used to filter out the undesired effects from the measured values.
In previous studies, we have therefore defined new metrics that quantify the intensity of coupling among layers of an undirected multiplex network, introducing the concept of \emph{multiplexity}~\cite{gemmetto2015}.
We have used these metrics to extensively document the empirical properties of real-world systems such as the World Trade Multiplex (WTM)~\cite{barigozzi2010,mastrandrea2014,gemmetto2015} and the European Airport Multiplex~\cite{cardillo2013}.
We concluded that much of the apparent multiplexity observed among the layers is actually explained by the local properties of nodes. 
Still, we found a significant level of measured remaining overlap, which quantifies the residual, `genuine' multiplexity structure of the WTM.

Whenever it is important to take into account the directionality of the connections in a graph~\cite{ebel2002}, the aforementioned approach can be extended to directed multi-layer networks~\cite{gemmetto2016}. 
We found that, in the directed case, the inter-layer `link overlap' can manifest itself in terms of both the `alignment' (a phenomenon that we called \emph{multiplexity} in analogy with the undirected case~\cite{gemmetto2015}) and the `anti-alignment' (a phenomenon that we called \emph{multireciprocity} as a generalization of the ordinary reciprocity for single-layer networks~\cite{garlaschelli2004,squartini2013}) of links across layers.
Since in each layer links are allowed in both directions between any two nodes, the alignment and the anti-alignment of links across layers do not conflict with each other and can actually coexist.

\subsection{A multiplex model with dyadic independence}
Our aim is that of introducing a minimal but realistic multiplex model that can reproduce the observed inter-layer dependencies reported above. 
Unlike the null models considered therein, the multiplex model should be characterized by nontrivial joint probabilities of connection involving multiple layers.
We want to develop one such model for binary multiplexes, and one for weighted multiplexes, in both the undirected and directed case. 

To keep the model as simple as possible, we assume \emph{dyadic independence}: the presence (and weight) of a link connecting a pair of nodes in a given layer does not depend on the presence (and weight) of a link connecting a \emph{different} pair of nodes in the same or in any other layer, although it does depend on the presence (and weight) of the links connecting the same pair of nodes in other layers. 
If we introduce the term \emph{multidyad} to denote a single pair of nodes `replicated' over all layers of the multiplex (i.e. the set of all single-layer dyads involving the same two nodes), the above assumption might be referred to as \emph{multidyadic independence}.
Note that, in directed and binary single-layer networks, a dyad formed by two nodes $i$ and $j$ can have 4 different topologies (a single link from $i$ to $j$, a single link from $i$ to $j$, two reciprocal links between $i$ and $j$, or no link at all). 
This implies that, in a directed binary multiplex with $M$ layers, a multidyad can have $4^M$ possible topologies. 
In a directed and weighted single-layer network, even assuming that the weights are non-negative integer numbers (as often done in previous approaches), a dyad can already have an infinity of possible weight-dependent configurations. Correspondingly, a multidyad in a multiplex with $M$ layers would have an infinite number, `raised to the $M$th power', of configurations.
Analogously, similar considerations hold for the undirected case, with the only difference that a dyad in a single-layer unweighted graph can now have 2 possible distinct values (a link between $i$ and $j$, or no link at all).

The assumption of multidyadic independence only restricts the topological properties that individual layers can have, but does not restrict the range of possible dependencies among layers of the multiplex.
Moreover, many single-layer networks have been in fact shown to have a structure consistent with dyadic independence~\cite{squartini2011,squartini2011I,squartini2011II,mastrandrea2014}. This property is also confirmed by the success of network reconstruction techniques that, as the one we will introduce here, assume dyadic independence~\cite{musmeci2013,caldarelli2013,mazzarisi2017,anand2017}.
An important example is given precisely by the WTM, whose single-layer structure is largely consistent with dyadic independence~\cite{squartini2011I,mastrandrea2014}.

\section{Binary multiplex model\label{sec:binary}}

In this Section, we develop our analytical framework and show the results of the application of such a theoretical model to a real-world system, namely the binarized version of the International Trade Multiplex~\cite{Baci,gemmetto2015}.  

Let us consider the marginal - i.e. unconditional on the presence of any other link in any layer - probability that a (possibly directed) link from node $i$ to node $j$ exists in layer $\alpha$:
\begin{equation}
p_{ij}^\alpha\equiv P(a_{ij}^\alpha=1)=\langle a_{ij}^\alpha\rangle
\end{equation}
(here and in what follows, angular brackets do not denote expected values under a \emph{null} model with \emph{independent} layers -- as for instance in~\cite{gemmetto2015} and~\cite{gemmetto2016} -- but ensemble averages over a \emph{realistic} multiplex model with \emph{dependent} layers). Due to our assumption of multidyadic independence, the relevant information that is marginalized in the  probability $p_{ij}^\alpha$ does not involve other pairs of nodes (joint probabilities involving multiple pairs of nodes would in any case factorize into products of marginal probabilities of invididual pairs of nodes), but it does involve other layers.

In other words, $p_{ij}^\alpha$ does not contain information about the inter-layer dependencies that we want to model. As such, it can be chosen to be specified by any convenient single-layer network model that satisfactorily reproduces the topological properties of layer $\alpha$. This marginal model is not actually an essential ingredient of our multiplex model and can be in some sense `outsourced'. For instance, it can be chosen to be a proper null model: an appropriate choice would be the (undirected or directed) Configuration Model~\cite{maslov2002}, i.e. the ensemble of networks satisfying on average the empirical degree sequence observed in that specific layer $\alpha$. It has indeed been shown~\cite{squartini2011,squartini2011I,mastrandrea2014} that this model is able to reliably replicate the topological properties of each layer of many real multilayer networks, including the World Trade Web itself~\cite{fagiolo2008,fagiolo2010}. Hence, defining the values $p_{ij}^\alpha$ as the link probabilities, for each layer separately, deriving from the Configuration Model is the most straightforward choice.
As a byproduct, this choice illustrates that the previously introduced multiplex assortativity metrics (Eq. (\ref{assort_old})) are not informative about the inter-layer coupling of interest for our analysis, because they are completely reabsorbed into the dyadic probabilities $p_{ij}^\alpha$; hence, these measures simply refer to a different kind of dependency between layers.

We now come to the definition of the true building blocks of our model of multiplexes with dependent layers.
Indeed, the assumption that layers are dependent implies that joint probabilities involving the same pairs of nodes but different layers should \emph{not} trivially factorize into products of marginal probabilities of the type $p_{ij}^\alpha$.
We therefore need to introduce generic joint probabilities that involve multiple layers. 
In general, even if we are assuming multidyadic independence, for each pair of nodes we should consider the joint probabilities of all combinations of links across all layers together, i.e. (in the jargon of multiplex networks~\cite{bianconi2013}) the probabilities of all possible \emph{multilinks} involving the same two nodes. 
As we mentioned, in multiplexes with directed links a multidyad can have $4^M$ possible topologies, i.e. $4^M$ possible multilinks. 
For each pair of nodes, fully specifying the joint connection probabilities across all layers would require the specification of a different probability for each of these multilinks, with the only constraint that the $4^M$ probabilities sum up to one.
This would lead to the definition of $4^M-1$ probabilities.
While this operation is feasible and insightful in the most studied case of a multiplex with two layers only, it becomes increasingly challenging (and decreasingly transparent) as $M$ increases. 

By contrast, we want to keep our approach feasible and useful (both from a modelling and from a network reconstruction perspective) even in the case of a very large number of layers, for which our formalism based on multiplexity and multireciprocity matrices fully shows its advantages.
Therefore we take the following parsimonious approach. For a given pair of nodes, we start from the definition of two joint (and conditional) probabilities that fully characterize both the multiplexity and the reciprocity properties of a single pair of layers, and then consider the set of such probabilities for all the $M^2$ pairs of layers (including a layer with itself) of the multiplex. This leads to a set of only $2 M^2$ probabilities defining the directed multiplex model (but still for a single pair of nodes).
This set represents the relevant projection (or marginalization) of the full set of $4^M-1$ multilink probabilities. The quadratic (as opposed to exponential) growth of the number of probabilities with the number of layers makes our approach appealing and manageable. 
Moreover, we will show that, at least in the empirical case study considered here, the conditional probabilities are approximately independent of the particular pair of nodes, making the information contained in the multiplexity and multireciprocity matrices sufficient in order to fully characterize the dependencies among the layers. 
Remarkably, this also means that the number of relevant probabilities remains $2 M^2$ (equal to the total number of entries in the multiplexity and multireciprocity matrices) independently of the number $N$ of nodes in the multiplex. Similar considerations can be made for the undirected systems.

We recall that, as reported in~\cite{gemmetto2015}, in the undirected binary case the multiplexity reads:
\begin{equation}
m_b^{\alpha \beta}= \frac{2 \sum_{i} \sum_{j < i} \min \{ a_{ij}^{\alpha}, a_{ij}^{\beta}\}}{L^{\alpha} + L^{\beta}}= \frac{2 \sum_{i} \sum_{j < i}  a_{ij}^{\alpha} a_{ij}^{\beta}}{L^{\alpha} + L^{\beta}} = \frac{2 L^{\alpha\rightrightarrows\beta}}{L^{\alpha} + L^{\beta}}
\label{mul_und_bin}
\end{equation}
where $ a_{ij}^{\alpha} $ are the entries of the adjacency matrices of the various layers, $ L^{\alpha} = \sum_{i < j} a_{ij}^{\alpha} $ is the number of links in that layer and $ L^{\alpha\rightrightarrows\beta} $ counts the number of links present in both layers $ \alpha $ and $ \beta $ between the same pairs of nodes. This notation is somewhat redundant at this stage, but on the other hand it allows for an easier generalization to the directed case, as we will show later. So, $ m_b^{\alpha \beta} $ ranges between 0 and 1 and represents a normalized overlap between pairs of layers of a multiplex. As mentioned in the Introduction, in the directed case we must take into account both the 'aligned' and the 'anti-aligned' overlap. Hence, in~\cite{gemmetto2016} we defined the binary directed multiplexity and multireciprocity respectively as:
\begin{equation}
m_b^{\alpha \beta}= \frac{2 \sum_{i} \sum_{j \neq i} \min \{ a_{ij}^{\alpha}, a_{ij}^{\beta}\}}{L^{\alpha} + L^{\beta}}= \frac{2 \sum_{i} \sum_{j \neq i} a_{ij}^{\alpha} a_{ij}^{\beta}}{L^{\alpha} + L^{\beta}} = \frac{2 L^{\alpha\rightrightarrows\beta}}{L^{\alpha} + L^{\beta}}
\label{mul_dir_bin}
\end{equation}
and
\begin{equation}
r_b^{\alpha \beta}= \frac{2 \sum_{i} \sum_{j \neq i} \min \{ a_{ij}^{\alpha}, a_{ji}^{\beta}\}}{L^{\alpha} + L^{\beta}}= \frac{2 L^{\alpha\leftrightarrows\beta}}{L^{\alpha} + L^{\beta}}
\label{rec_dir_bin}
\end{equation}
where $ L^{\alpha\rightrightarrows\beta} $ represents the number of directed links present in both the considered layers between the same pairs of nodes, while $ L^{\alpha\leftrightarrows\beta} $ counts the number of directed links present in $ \alpha $ which are reciprocated in $ \beta $, over all the possible pairs of vertices.

In the previous sections we stressed the importance of the inter-layer link coupling for the characterization of a real-world multiplex. We now pave the way for realistic (undirected and directed) binary models that can capture the observed features in the particular case of the World Trade Multiplex. Once more, one should not confuse these realistic models with the null models used in other contexts~\cite{park2004}.

\subsection{Undirected binary model}

We start with the definitions of the measures we will focus on in our analysis. The empirical single-layer degree reads: 
\begin{eqnarray}
k_i^{\alpha}  = \sum_{j \neq i} a_{ij}^{\alpha}.
\label{emp_deg_und}
\end{eqnarray}
Moreover, we can introduce the first of the new quantities that will allow us to properly describe the inter-layer coupling of a multiplex, namely the empirical \emph{multiplexed degree}:
\begin{equation}
k_i^{\alpha \rightrightarrows \beta} = \sum_{j \neq i} a_{ij}^{\alpha} a_{ij}^{\beta}.
\label{emp_multipl_deg_und}
\end{equation}
If we look at Eq.~(\ref{mul_und_bin}), we immediately see that, as compared to the global quantity $m_b^{\alpha,\beta}$, the multiplexed degree $k_i^{\alpha \rightrightarrows \beta}$ provides an even more detailed, local quantification of the multiplexity.

\begin{figure*}[ht]
\begin{center}
\includegraphics[width=1.0\textwidth]{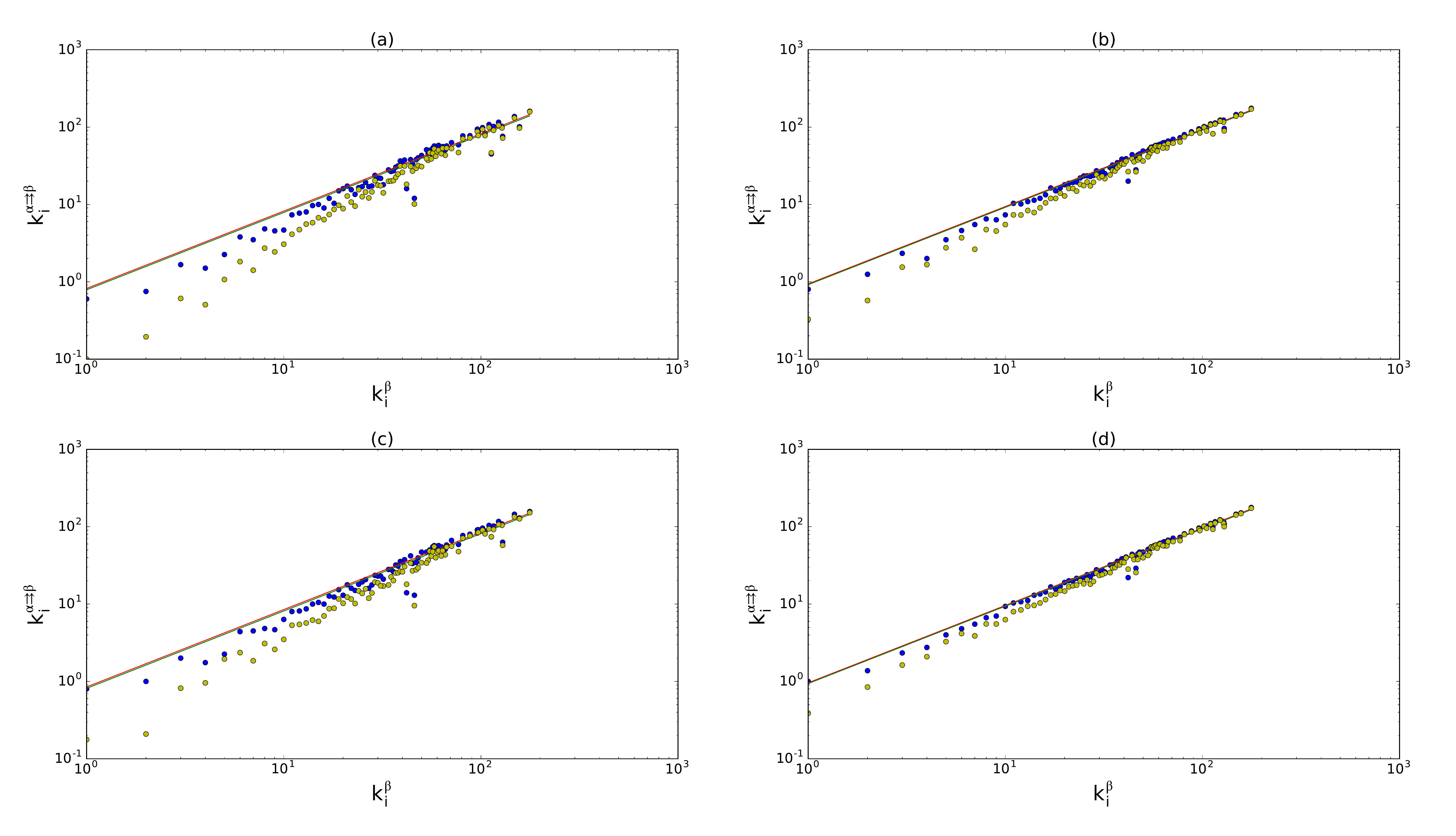}
\end{center}
\caption{Degree of layer $ \beta $ versus inter-layer multiplexed degree for 4 different pairs of commodities: inorganic chemicals (a), plastics (b), iron and steel (c), electric machinery (d)
versus trade in cereals. Blue dots: real data; yellow dots: expected multiplexed degree according to the uncorrelated model; lower green line: expected trend according to (\ref{rel_r_rbin}); upper red line (when discernible): best fit. In all the cases, $ R^2 > 0.93 $, for both the curves. It should be noted that we fit the empirical data with lines of the form $ y = a \cdot x $, and only after we plot the results in log-log scale.}
\label{fig:degree_mul_und}
\end{figure*}

In what follows, we first establish an empirically robust pattern displayed by $k_i^{\alpha \rightrightarrows \beta}$ and then select it as one of the target properties that a multiplex reconstruction model should replicate, in addition to the desired marginal single-layer network properties.  
Figure~\ref{fig:degree_mul_und} reports the scatter plot of $ k_i^{\alpha \rightrightarrows \beta} $ versus $ k_i^{\beta} $ for four pairs of commodities (blue points). We clearly see an approximate linear trend of the type
\begin{equation}
k_i^{\alpha \rightrightarrows \beta} \approx u^{\alpha \beta} k_i^{\beta}.
\label{lin_emp}
\end{equation}
Similar plots can be observed for the other pairs of layers as well (not shown). The robustness of this pattern motivates us to look for a multiplex model able to replicate it.

We define the joint probability $p^{\alpha \rightrightarrows \beta} _{ij}$ for the simultaneous presence of a link from node $i$ to node $j$ in layer $\alpha$ and of a corresponding link in layer $\beta$:
\begin{equation}
p_{ij}^{\alpha \rightrightarrows \beta}\equiv 
P(a^\alpha_{ij}=1\cap a^\beta_{ij}=1)=
\langle a_{ij}^{\alpha} a_{ij}^{\beta} \rangle = 
p_{ij}^{\beta \rightrightarrows \alpha}.
\end{equation}
Using $p_{ij}^{\alpha \rightrightarrows \beta}$ and the aforementioned $p_{ij}^{\beta}$ we can also obtain the \emph{conditional} probability $u_{ij}^{\alpha \beta}$ that a link from $i$ to $j$ exists in layer $\alpha$, given that the corresponding link exists in layer $\beta$:
\begin{equation}
u_{ij}^{\alpha \beta}\equiv 
P(a^\alpha_{ij}=1 | a^\beta_{ij}=1)=
p_{ij}^{\alpha \rightrightarrows \beta}/
p_{ij}^{\beta }.
\end{equation}
We call $u_{ij}^{\alpha \beta}$ the \emph{multiplexity probability}.
Note that, while $p_{ij}^{\alpha \rightrightarrows \beta}$ is symmetric under the exchange of $\alpha$ and $\beta$, $u_{ij}^{\alpha \beta}$ is not;  indeed, we have:
\begin{equation}
p_{ij}^{\alpha \rightrightarrows \beta}=u_{ij}^{\alpha \beta} p_{ij}^{\beta }=u_{ij}^{\beta \alpha}p_{ij}^{\alpha} = p_{ij}^{\beta \rightrightarrows \alpha}.
\end{equation}
Furthermore $ p_{ij}^{\alpha \rightrightarrows \beta} $ depends, at least in the general case, both on the pair of nodes and on the pair of layers. 
Given the previous definitions, the expected value of the multiplexed degree becomes:
\begin{eqnarray}
\langle k_i^{\alpha \rightrightarrows \beta}\rangle = \sum_{j \neq i} \langle a_{ij}^{\alpha} a_{ij}^{\beta} \rangle=\sum_{j \neq i}p_{ij}^{\alpha \rightrightarrows \beta}= \sum_{j \neq i}u_{ij}^{\alpha \beta}p_{ij}^{\beta}=\sum_{j \neq i}u_{ij}^{\beta \alpha}p_{ij}^{\alpha}.
\label{eq:expmultip}
\end{eqnarray}
The main goal consists in understanding the structure of $u_{ij}^{\alpha \beta}$, which is the crucial quantity responsible for the coupling among layers. By contrast, as already said before, $p_{ij}^{\beta}$ can in general be left largely unspecified as it can be chosen to be any single-layer network model that satifactorily reproduces a set of desired marginal topological properties of layer $\beta$, irrespective of the coupling with the other layers. 
The only basic property we require from $p_{ij}^{\beta}$ is that the degree sequence is among such desired properties, or in other words that, for each node $i$ and each layer $\beta$, the expected degree $\langle k_i^\beta\rangle=\sum_{j\ne i}p_{ij}^\beta$ satisfactorily replicates the empirical degree $k_i^\beta$:
\begin{equation}
\langle k_i^\beta\rangle=\sum_{j\ne i}p_{ij}^\beta\approx k_i^\beta\quad\forall i.
\label{eq:matchk}
\end{equation}
For instance, if the Binary Configuration Model~\cite{squartini2011,squartini2015} is chosen as the marginal single-layer reconstruction method, the above criterion is strictly verified, since that model assumes that the degree of each node is known and  that the $p_{ij}^\beta$ can be \emph{constructed} as the maximum-entropy probability such that
\begin{equation}
\langle k_i^\beta\rangle=\sum_{j\ne i}p_{ij}^\beta= k_i^\beta\quad\forall i.
\label{eq:matchkCM}
\end{equation}
Other marginal reconstruction methods, which relax the hypothesis that the degree of each node is known, use other node-specific pieces of information, plus some proxy of the overall network density, to construct a $p_{ij}^\beta$ such that Eq.~(\ref{eq:matchk}) is in any case realized~\cite{garlaschelli2004a,garlaschelli2005,cimini1,cimini2,cimini3}.
The above examples have all been shown to provide reliably reconstructed networks~\cite{cimini1,cimini2,cimini3}.

The presence of a nontrivial $u_{ij}^{\alpha \beta}$ in the present multiplex model implies that any $p_{ij}^{\beta}$ coming from a single-layer model should be interpreted as a marginal probability resulting from a more realistic model where the presence of links across all layers is governed by a joint distribution for the entire multiplex.
In other words, $u_{ij}^{\alpha \beta}$ allows us to extend any desired single-layer model to a truly multiplex model with nontrivial coupling among layers.
The trivial case of independent layers can be easily recovered by setting:
\begin{equation}
\left[ u_{ij}^{\alpha \beta} \right]_{unc} =p_{ij}^\alpha
\label{prob_uncorr}
\end{equation}
since here the presence of the link in layer $\beta$ does not affect the connection probability in layer $\alpha$. In such a case, the expected multiplexed degree becomes:
\begin{equation}
\langle k_i^{\alpha \rightrightarrows \beta}\rangle_{unc} = \sum_{j \neq i} p_{ij}^\alpha p_{ij}^\beta.
\label{mul_deg_uncorr}
\end{equation}
From Eq.~(\ref{mul_und_bin}) it should be noted that, if such an uncoupled model were used to generate the multiplex, the expected value of the multiplexity $m_b^{\alpha\beta}$ would be zero.
Yet, if Eq.~(\ref{eq:matchk}) holds, then the inter-layer degree correlation function defined in Eq.~(\ref{assort_old}) would be replicated. This shows that such a correlation function is not informative about the genuine inter-layer dependencies which go beyond the degree-degree correlations across the layers of the multiplex.
By contrast, the multiplexity $m_b^{\alpha\beta}$ is, confirming the argument that led us to its introduction in ref.~\cite{gemmetto2015}.

To build a minimal model that can reproduce the observed level of similarity (i.e., multiplexity) between layers of the multiplex, we require that the robust empirical trend encapsulated in Eq.~(\ref{lin_emp}) is replicated.
Looking at Eqs.~(\ref{eq:expmultip}) and~(\ref{eq:matchk}), and imposing Eq.~(\ref{lin_emp}), this requirement implies that the conditional probability $u_{ij}^{\alpha \beta}$ should be approximately independent of the pair of nodes:
\begin{eqnarray}
u_{ij}^{\alpha \beta} = \frac{p_{ij}^{\alpha \rightrightarrows \beta}}{p_{ij}^{\beta}} = \frac{\langle a_{ij}^{\alpha} a_{ij}^{\beta} \rangle}{\langle a_{ij}^{\beta} \rangle} \approx u^{\alpha \beta}.
\label{def_rcond}
\end{eqnarray}
Since the transformation $ i \mapsto j $ together with $ \alpha \mapsto \beta $ keeps the quantities unaffected, we also have
\begin{eqnarray}
u^{\alpha \beta} \langle a_{ij}^{\beta} \rangle \approx \langle a_{ij}^{\alpha} a_{ij}^{\beta} \rangle = \langle a_{ij}^{\beta} a_{ij}^{\alpha} \rangle \approx q^{\beta \alpha} \langle a_{ij}^{\alpha} \rangle.
\label{star}
\end{eqnarray}
Summing over $ i $ and $ j $, we get
\begin{eqnarray}
u^{\alpha \beta} L^{\beta}\approx u^{\beta \alpha} L^{\alpha}.
\label{global_L}
\end{eqnarray}
and from (\ref{def_rcond}) we immediately have
\begin{eqnarray}
u^{\alpha \beta} \langle a_{ij}^{\beta} \rangle \approx \langle a_{ij}^{\alpha} a_{ij}^{\beta} \rangle.
\label{def_rcond_inv}
\end{eqnarray}
Summing over $i$ and $j$ and inverting, we obtain
\begin{eqnarray}
u^{\alpha \beta} \approx \frac{\sum_i\sum_{i < j}\langle a_{ij}^{\alpha} a_{ij}^{\beta} \rangle }{\sum_i\sum_{i < j} \langle a_{ij}^{\beta} \rangle } = \frac{\sum_i\sum_{i < j}a_{ij}^{\alpha} a_{ij}^{\beta} }{L^{\beta} }.
\end{eqnarray}
The above relations allow us to express twice the inverse of (\ref{mul_und_bin}) as
\begin{eqnarray}
\frac{2}{m_b^{\alpha \beta}} = \frac{ L^{\alpha} + L^{\beta}  }{\sum_i\sum_{i < j} a_{ij}^{\alpha} a_{ij}^{\beta} } \approx \frac{1}{u^{\alpha\beta}} + \frac{1}{u^{\beta \alpha}}
\end{eqnarray}
where $ m_b^{\alpha \beta} $ is measured from the multiplex data while $ u^{\alpha \beta} $ is derived from the slope of the empirical linear relationship between $ k_i^{\beta} $ and $ k_i^{\alpha \rightrightarrows \beta} $.
Thus, we find that $ m_b^{\alpha \beta} $ is approximately the harmonic mean of the conditional probabilities $ u^{\alpha \beta} $ and $ u^{\beta \alpha} $.
Applying Eq.~(\ref{global_L}) to the previous expression, we get:
\begin{eqnarray}
\frac{2}{m_b^{\alpha \beta}} & \approx & \frac{1}{u^{\alpha \beta}} \left( 1 + \frac{u^{\alpha \beta}}{u^{\beta \alpha}} \right)  \nonumber \\
                                 & \approx & \frac{1}{u^{\alpha \beta}} \left( 1 + \frac{L^{\alpha}}{L^{\beta}} \right)  \nonumber \\
                                 & = & \frac{L^{\alpha} + L^{\beta}}{u^{\alpha \beta} L^{\beta}}
\end{eqnarray}
Hence, the value of the slope in the plots of $ k_i^{\alpha \rightrightarrows \beta} $ vs $ k_i^{\beta} $ is predicted to be
\begin{eqnarray}
u^{\alpha \beta} \approx \frac{L^{\alpha} + L^{\beta}}{2 L^{\beta}} m_b^{\alpha \beta}
\label{rel_r_rbin}
\end{eqnarray}

Indeed, in Figure~\ref{fig:degree_mul_und} we show that the best fit curves almost coincide with the expected ones having slope calculated independently from Eq.~(\ref{rel_r_rbin}). Futhermore, we also show (yellow dots) that the model assuming independent layers as in Eqs.~(\ref{prob_uncorr}) and (\ref{mul_deg_uncorr}) produces values of the multiplexed degree that are systematically lower than the empirical ones.

From the previous analysis, it turns out phenomenologically that the minimal model one can design in order to reproduce the (local) observed values of the multiplexed degree requires only the (global)
information about the total number of multiplexed links 
$ L^{\alpha \rightrightarrows \beta} $ for any ordered pair of layers $ (\alpha,\beta) $ (together with the aforementioned degree sequences in each layer).

In other words, a reliable network reconstruction method for the class of multiplexes we are focusing on here requires as input information a reconstruction model that works successfully on each layer separately, plus the $ M (M-1)/2 $ values of $ L^{\alpha \rightrightarrows \beta} $, for all pairs of layers. These values are the numerators of the entries of the so-called (binary) \emph{multiplexity matrix} \cite{gemmetto2015}.
If the reconstruction model is chosen to be the Configuration Model, then the overall input information reduces to the degree sequence $\vec{k}^\alpha$ for each layer $\alpha$, plus the values $ L^{\alpha \rightrightarrows \beta} $ for each pair of layers.

\subsection{Directed binary model}

As said in the introductive section, in the directed case we should take into account that the inter-layer coupling can intervene both in terms of alignment and anti-alignment. Hence, we have not only to extend the notion of \emph{multiplexed degree} to the directed case, but also to introduce the quantity dubbed \emph{multireciprocated degree}. It is indeed straightforward to exploit the same approach to analyse the patterns of multiplexity and multireciprocity in the directed case. The main difference w.r.t. the undirected case will consist in the definition of two separate conditional probabilities.
We start defining the quantities that we will measure on the real multiplex network, namely the in-degree:
\begin{eqnarray}
k_i^{\alpha,in}  = \sum_{j \neq i} a_{ji}^{\alpha};
\label{emp_indeg_und}
\end{eqnarray}
and the out-degree:
\begin{eqnarray}
k_i^{\alpha,out}  = \sum_{j \neq i} a_{ij}^{\alpha}.
\label{emp_outdeg_und}
\end{eqnarray}
In analogy with the undirected model, we assume we can start from a marginal single-layer model characterized by the probability $p_{ij}^\beta=\langle a_{ij}^\beta\rangle$ that a \emph{directed} link from node $i$ to node $j$ exists.
The only thing we require from $p_{ij}^\beta$ is that it reliably replicates the in- and out-degree of each node $i$ in layer $\beta$:
\begin{eqnarray}
\langle k_i^{\alpha,in}\rangle&=&\sum_{j\ne i}p_{ji}^\alpha\approx k_i^{\alpha,in}\quad\forall i\label{eq:matchkin}\\
\langle k_i^{\alpha,out}\rangle&=&\sum_{j\ne i}p_{ij}^\alpha\approx k_i^{\alpha,out}\quad\forall i,
\label{eq:matchkout}
\end{eqnarray}
generalizing the corresponding criterion in Eq.~(\ref{eq:matchk}).

\begin{figure*}[htbp]
\begin{center}
\includegraphics[width=1.0\textwidth]{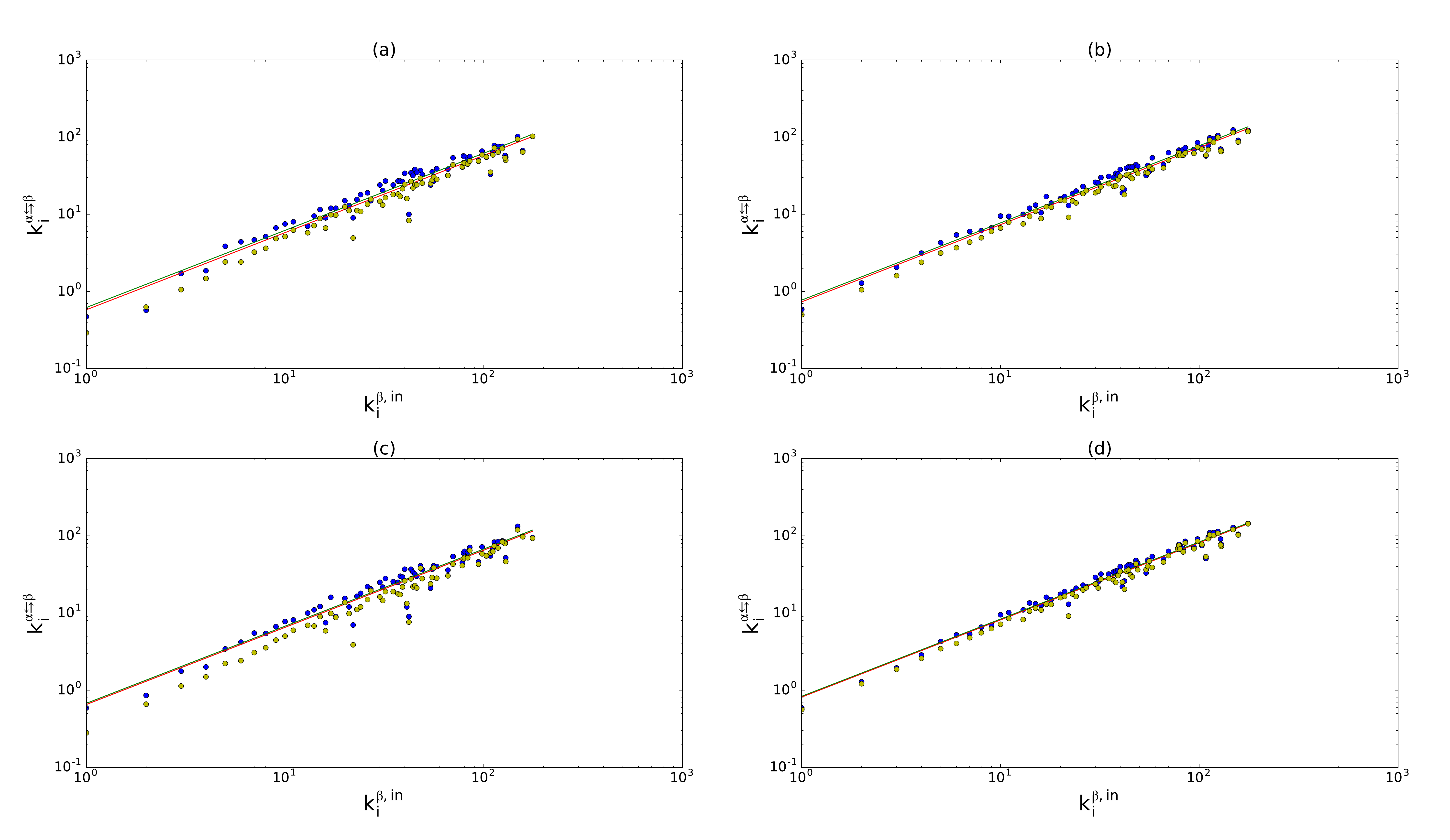}
\end{center}
\caption{In-degree of layer $ \beta $ versus inter-layer multireciprocated degree for 4 different pairs of commodities: inorganic chemicals (a), plastics (b), iron and steel (c), electric machinery (d)
versus trade in cereals. Blue dots: real data; yellow dots: expected multireciprocated degree according to the uncorrelated model; lower green line: expected trend according to (\ref{rel_v_r_dir}); upper red line (when discernible): best fit. In all the cases, $ R^2 > 0.95 $, for both the curves. It should be noted that we fit the empirical data with lines of the form $ y = a \cdot x $, and only after we plot the results in log-log scale.}
\label{fig:degree_rec_dir}
\end{figure*}

We also define the multiplex quantities that extend the ones introduced in the undirected case, i.e. the \emph{multiplexed degree}:
\begin{equation}
k_i^{\alpha \rightrightarrows \beta} = \sum_{j \neq i} a_{ij}^{\alpha} a_{ij}^{\beta}.
\label{emp_multipl_deg_dir}
\end{equation}
and the \emph{multireciprocated degree}.
\begin{equation}
k_i^{\alpha \rightleftarrows \beta} = \sum_{j \neq i} a_{ij}^{\alpha} a_{ji}^{\beta}.
\label{emp_multirec_deg_dir}
\end{equation}
It is possible to generalize the argument explained in the previous subsection; also in this case we find that $ k_i^{\alpha \rightrightarrows \beta} $ and $ k_i^{\alpha \rightleftarrows \beta} $ are in almost-linear relation with, respectively, $ k_i^{\beta,out} $ (not shown, as it is very similar to the undirected case) and $ k_i^{\beta,in} $ (Figure~\ref{fig:degree_rec_dir}, blue dots), therefore we can set:
\begin{equation}
k_i^{\alpha \rightrightarrows \beta} \approx u^{\alpha \beta} k_i^{\beta,out}
\label{lin_emp_mul_dir}
\end{equation}
and 
\begin{equation}
k_i^{\alpha \rightleftarrows \beta} \approx v^{\alpha \beta} k_i^{\beta,in}.
\label{lin_emp_rec_dir}
\end{equation}
The presence of two different multiplex quantities leads to the definition of two distinct joint probabilities: 
\begin{equation}
p_{ij}^{\alpha \rightrightarrows \beta}\equiv 
P(a^\alpha_{ij}=1\cap a^\beta_{ij}=1)=
\langle a_{ij}^{\alpha} a_{ij}^{\beta} \rangle = 
p_{ij}^{\beta \rightrightarrows \alpha}
\end{equation}
gives the probability for the simultaneous presence of a link from node $i$ to node $j$ in layer $\alpha$ and of a corresponding link (with the same direction) in layer $\beta$, while:
\begin{equation}
p_{ij}^{\alpha \rightleftarrows \beta}\equiv 
P(a^\alpha_{ij}=1\cap a^\beta_{ji}=1)=
\langle a_{ij}^{\alpha} a_{ji}^{\beta} \rangle = 
p_{ji}^{\beta \rightleftarrows \alpha}
\end{equation}
is the probability of having a link from node $i$ to node $j$ in layer $\alpha$ and a link in the opposite direction in layer $\beta$.
Consequently, from these joint probabilities and the marginal single-layer probabilities we can derive the two separate \emph{conditional} probability $u_{ij}^{\alpha \beta}$ that a link from $i$ to $j$ exists in layer $\alpha$, given that the corresponding link exists in layer $\beta$:
\begin{equation}
u_{ij}^{\alpha \beta}\equiv 
P(a^\alpha_{ij}=1 | a^\beta_{ij}=1)=
p_{ij}^{\alpha \rightrightarrows \beta}/
p_{ij}^{\beta }.
\end{equation}
and $v_{ij}^{\alpha \beta}$ representing the probability of having a link from $i$ to $j$ in $\alpha$, given that a link from $j$ to $i$ exists in layer $\beta$:
\begin{equation}
v_{ij}^{\alpha \beta}\equiv 
P(a^\alpha_{ij}=1 | a^\beta_{ji}=1)=
p_{ij}^{\alpha \rightleftarrows \beta}/
p_{ji}^{\beta }.
\end{equation}
We call $u_{ij}^{\alpha \beta}$ the \emph{multiplexity probability} and $v_{ij}^{\alpha \beta}$ the \emph{multireciprocity probability}.
These probabilities lead to the separate notions of expected \emph{multiplexed} and \emph{multireciprocated degree}, defined respectively as:
\begin{eqnarray}
\langle k_i^{\alpha \rightrightarrows \beta}\rangle = \sum_{j \neq i} \langle a_{ij}^{\alpha} a_{ij}^{\beta} \rangle=\sum_{j \neq i}p_{ij}^{\alpha \rightrightarrows \beta}= \sum_{j \neq i}u_{ij}^{\alpha \beta}p_{ij}^{\beta}=\sum_{j \neq i}u_{ij}^{\beta \alpha}p_{ij}^{\alpha} 
\label{multipl_deg_dir}
\end{eqnarray}
and:
\begin{eqnarray}
\langle k_i^{\alpha \rightleftarrows \beta}\rangle = \sum_{j \neq i} \langle a_{ij}^{\alpha} a_{ji}^{\beta} \rangle=\sum_{j \neq i}p_{ij}^{\alpha \rightleftarrows \beta}= \sum_{j \neq i}v_{ij}^{\alpha \beta}p_{ji}^{\beta}=\sum_{j \neq i}v_{ij}^{\beta \alpha}p_{ji}^{\alpha}
\label{multirec_deg_dir}
\end{eqnarray}
Analogously to the undirected case, $u_{ij}^{\alpha \beta}$ and $v_{ij}^{\alpha \beta}$ are driving the real coupling among the layers of the system, while the single-layer probabilities $ p_{ij}^{\alpha} $ can be freely chosen starting from any network model that correctly reproduces the marginal topology of the considered layer.
For instance, we may choose the Directed Configuration Model~\cite{squartini2011,squartini2015}, for which Eqs.~(\ref{eq:matchkin}) and~(\ref{eq:matchkout}) hold with a strict equality sign, or some of its relaxed versions that assume less input information~\cite{cimini1,cimini2,cimini3}.

With the same reasoning of the previous subsection, it is possible to show that the value of the slope in the plots of $ k_i^{\alpha \rightrightarrows \beta} $ vs $ k_i^{\beta,out} $ is predicted to be:
\begin{eqnarray}
u^{\alpha \beta} \approx \frac{L^{\alpha} + L^{\beta}}{2 L^{\beta}} m_b^{\alpha \beta}
\label{rel_u_m_dir}
\end{eqnarray}
while the slope in the plots of $ k_i^{\alpha \rightleftarrows \beta} $ vs $ k_i^{\beta,in} $ is, according to the model:
\begin{eqnarray}
v^{\alpha \beta} \approx \frac{L^{\alpha} + L^{\beta}}{2 L^{\beta}} r_b^{\alpha \beta}.
\label{rel_v_r_dir}
\end{eqnarray}
As shown in Figure~\ref{fig:degree_rec_dir} for the multireciprocated degree (the corresponding plot referred to the multiplexed degree is not reported, being however very similar to the undirected case), the best fit curves are well modelled by the expected ones. We also show the results of the uncorrelated model, producing again values of the multireciprocated degree which are systematically lower than the observed values.

It turns therefore out that an appropriate multiplex reconstruction method for the class of directed multi-layer networks we are considering is based on the information about the in- and out-degree sequences of each layer combined with the entries of the matrices $ L^{\alpha \rightrightarrows \beta} $ and $ L^{\alpha \rightleftarrows \beta} $ for any pair of layers.

\section{Weighted multiplex model\label{sec:weighted}}

In the case of weighted multiplex networks, the marginal (i.e. single-layer) quantity we will focus on is the the weight $ w_{ij}^\alpha $ associated to any (possible directed) link between $ i $ and $ j $ in layer $ \alpha $, together with its expected value $ \langle w_{ij}^\alpha \rangle $. At the same time, we can still consider the link probability $p_{ij}^\alpha$, representing the chance that nodes $i$ and $j$ are connected by a link, irrespective of the weight of the latter. Since the assumption of multidyadic independence still holds, the information provided by $ \langle w_{ij}^\alpha \rangle $ and $p_{ij}^\alpha$ does not involve other pairs of nodes other than $ (i,j) $. 

As the marginal quantities $ \langle w_{ij}^\alpha \rangle $ are not influenced by the inter-layer coupling that we will add, they can therefore be considered as expectation values provided by any model able to correctly reproduce the weighted structure of layer $ \alpha $. However, in order to correctly reproduce the entire multiplex, we need to employ a single-layer model which has been proved to be reliable; it has been shown~\cite{mastrandrea2014bis} that the Weighted Configuration Model~\cite{serrano2005} is not capable of reproducing both the topology and the weighted structure of a network, as it gives rise to almost complete graphs. Instead, we can think of the marginal values as stemming from the Enhanced Configuration Model~\cite{mastrandrea2014,squartini2015} - constraining both the degree and the strength sequence of the observed graph, i.e.
\begin{eqnarray}
\langle s_i^\alpha\rangle&=&\sum_{j\ne i}\langle w^\alpha_{ij}\rangle=s_i^\alpha\quad\forall i,\label{eq:matchs}\\
\langle k_i^\alpha\rangle&=&\sum_{j\ne i}p^\alpha_{ij}=k_i^\alpha\quad\forall i.\label{eq:matchsk}
\end{eqnarray}
Similar to the binary case, these constraints can be relaxed in such a way that the required input information is considerably reduced. 
For instance, the methods proposed in refs.~\cite{cimini1,cimini2,cimini3} require much less input information but are still such that
\begin{eqnarray}
\langle s_i^\alpha\rangle&=&\sum_{j\ne i}\langle w^\alpha_{ij}\rangle\approx s_i^\alpha\quad\forall i,\label{eq:matchsapprox}\\
\langle k_i^\alpha\rangle&=&\sum_{j\ne i}p^\alpha_{ij}\approx k_i^\alpha\quad\forall i,\label{eq:matchskapprox}
\end{eqnarray}
and have recently been found to provide the best reconstruction methods for monoplex weighted networks from limited information~\cite{mazzarisi2017,anand2017}.

In the weighted case, the assumption of dependency between layers means that the joint probability of observing a given weight $ w_{ij}^\alpha $ between $ i $ and $ j $ in layer $ \alpha $ together with a weight $ w_{ij}^\beta $ in $ \beta $ does not factorize into two separate single-layer probabilities. In previous studies~\cite{menichetti2014} this issue has been tackled by introducing the concept of \emph{multistrength}; however, as already explained for the binary case, this approach is practically feasible only in the case of multiplex networks with a (very) limited number of layers.

On the contrary, our multiplex reconstruction technique appears to be useful also when applied to multigraphs possessing a larger number of layers, as it requires as input the strength sequence of the various layers and the multiplexity/multireciprocity matrices (both growing like $ M^2 $). This quadratic growth in the number of layers (opposed to the exponential growth shown by the multistrength method), combined with the phenomenological observation that the conditional probabilities are again independent of the considered pair of nodes, makes our approach very promising.
  
As we said, our reconstruction method builds on the notions of \emph{weighted multiplexity and multireciprocity}; in particular, in the undirected case we will exploit the measures of weighted multiplexity introduced in~\cite{gemmetto2015}:
\begin{equation}
m_w^{\alpha \beta}= \frac{2 \sum_i\sum_{j < i} \min \{ w_{ij}^{\alpha}, w_{ij}^{\beta}\}}{W^{\alpha} + W^{\beta}}= \frac{2 W^{\alpha\rightrightarrows\beta}}{W^{\alpha} + W^{\beta}}
\label{mul_und_wei}
\end{equation}
where $ w_{ij}^{\alpha} $ are the entries of the weighted adjacency matrices of the various layers, $ W^{\alpha} = \sum_{i < j} w_{ij}^{\alpha} $ is the total weight associated to the links in that layer and $ W^{\alpha\rightrightarrows\beta} $ represents the "shared weight" between $ \alpha $ and $ \beta $. In analogy with the binary case, $ m_w^{\alpha \beta} $ ranges between 0 and 1 and represents a normalized weighted overlap between pairs of layers of the multi-graph. In the directed case, instead, we have to consider the overlap in both the directions. In~\cite{gemmetto2016} we defined the weighted directed multiplexity and multireciprocity respectively as:
\begin{equation}
m_w^{\alpha \beta}= \frac{2 \sum_i\sum_{j \neq i} \min \{ w_{ij}^{\alpha}, w_{ij}^{\beta}\}}{W^{\alpha} + W^{\beta}}= \frac{2 W^{\alpha\rightrightarrows\beta}}{W^{\alpha} + W^{\beta}}
\label{mul_dir_wei}
\end{equation}
and
\begin{equation}
r_w^{\alpha \beta}= \frac{2 \sum_i\sum_{j \neq i} \min \{ w_{ij}^{\alpha}, w_{ji}^{\beta}\}}{W^{\alpha} + W^{\beta}}= \frac{2 W^{\alpha\rightleftarrows\beta}}{W^{\alpha} + W^{\beta}}
\label{rec_dir_wei}
\end{equation}
where $ W^{\alpha\rightrightarrows\beta} $ is the "shared total weight" between the considered layers, and $ W^{\alpha\leftrightarrows\beta} $ is the "shared reciprocated weight" between $ \alpha $ and $ \beta $.

In the following sections we will show a method to reconstruct the World Trade Multiplex from single-layer information exploiting the knowledge of the aforementioned multiplexity and multireciprocity matrices.

\subsection{Undirected weighted model}

\begin{figure*}[ht]
\begin{center}
\includegraphics[width=1.0\textwidth]{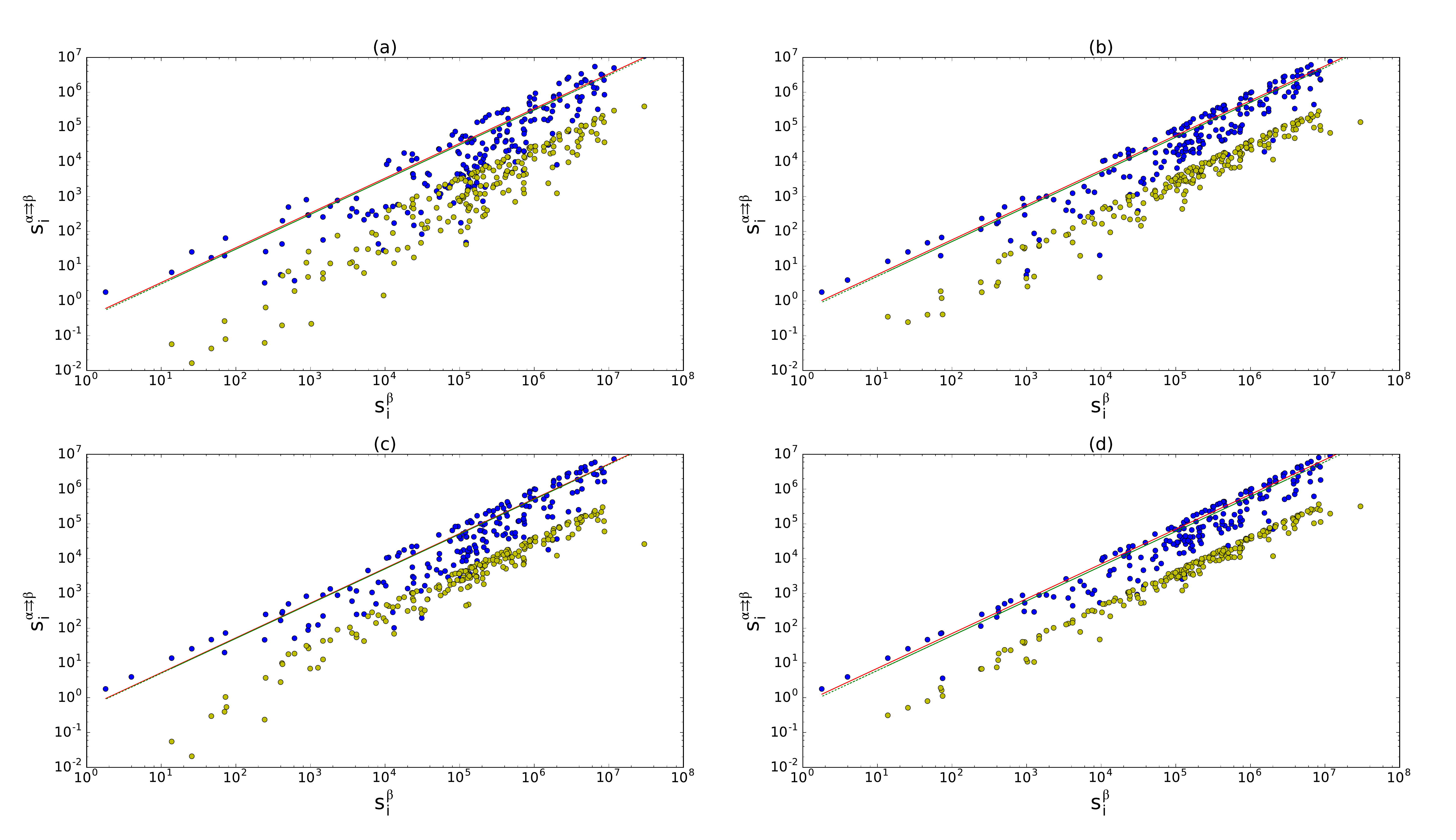}
\end{center}
\caption{Strength of layer $ \beta $ versus inter-layer multiplexed strength for 4 different pairs of commodities: inorganic chemicals (a), plastics (b), iron and steel (c), electric machinery (d)
versus trade in cereals. Blue dots: real data; yellow dots: expected multiplexed strength according to the uncorrelated model; lower green line: expected trend according to (\ref{rel_U_mul_und_wei}); upper red line (when discernible): best fit. In all the cases, $ R^2 > 0.92 $, for both the curves. It should be noted that we fit the empirical data with lines of the form $ y = a \cdot x $, and only after we plot the results in log-log scale.}
\label{fig:strength_mul_und}
\end{figure*}

In this section, we will focus on the relation between the single-layer strength, defined as:
\begin{eqnarray}
s_i^{\alpha} = \sum_{j \neq i}^{} w_{ij}^{\alpha}
\label{strength_seq_emp_und}
\end{eqnarray}
and the \emph{multiplexed strength}, for any ordered pair of layers:
\begin{eqnarray}
s_i^{\alpha \rightrightarrows \beta} = \sum_{j \neq i}^{}  w_{ij}^{\alpha \rightrightarrows \beta} \equiv \min \{ w_{ij}^{\alpha}, w_{ij}^{\beta} \}
\label{strength_seq_mul_emp_und}
\end{eqnarray}
where $ w_{ij}^{\alpha \rightrightarrows \beta} $ is the multiplexed component of the weights associated to the links between $ i $ and $ j $ in layers $ \alpha $ and $ \beta $.
In particular, $ s_i^{\alpha \rightrightarrows \beta} $ is the multiplex quantity allowing us to describe the inter-layer weighted coupling.
Figure~\ref{fig:strength_mul_und} reports the relation between $ s_i^{\beta} $ and $ s_i^{\alpha \rightrightarrows \beta} $ for various pairs of commodities of the World Trade Multiplex; a clear empirical trend is exhibited (blue points), that can be approximated as:
\begin{equation}
s_i^{\alpha \rightrightarrows \beta} \approx U^{\alpha \beta} s_i^{\beta}.
\label{lin_emp-wei}
\end{equation}
Our goal will consist in designing the minimal model able to capture this empirical evidence.

In this perspective, we define the corresponding expected quantities $ \langle w_{ij}^{\alpha \rightrightarrows \beta} \rangle $ and $ \langle w_{ij}^{\alpha} \rangle $; in particular, the multiplexed component can be written in terms of a joint probability, in order to keep the same structure adopted for the binary case:
\begin{eqnarray}
\langle w_{ij}^{\alpha \rightrightarrows \beta} \rangle & = &\! \langle \min \{ w_{ij}^{\alpha}, w_{ij}^{\beta} \} \rangle = \nonumber \\
                                                      & = &\! \sum_{w = 1}^{\infty} P \left( \min \{ w_{ij}^{\alpha}, w_{ij}^{\beta} \} \geq w \right) = \nonumber \\
                                                      & = &\! \sum_{w = 1}^{\infty} P \left( w_{ij}^{\alpha} \geq w \cap w_{ij}^{\beta} \geq w \right) = \nonumber \\
                                                      & = &\! \sum_{w = 1}^{\infty} \! U_{ij}^{\alpha \beta}\! \big( w_{ij}^{\alpha} \geq w | w_{ij}^{\beta} \geq w \big)   P \big( w_{ij}^{\beta} \geq w \big) 
\label{prob_cond_wei_und}
\end{eqnarray}
where $ U_{ij}^{\alpha \beta} $ is now the probability of observing a weight $ w_{ij}^{\alpha} $ in $ \alpha $ larger than $ w $ given that a weight $ w_{ij}^{\beta} $ larger than $ w $ has been observed in $ \beta $.

As mentioned, the phenomenological observation shows that the conditional probability defined in~(\ref{prob_cond_wei_und}) is actually independent from the considered pair of nodes:
\begin{eqnarray}
U_{ij}^{\alpha \beta} = \frac{\langle \min \{ w_{ij}^{\alpha}, w_{ij}^{\beta} \} \rangle }{\langle w_{ij}^{\beta} \rangle} \approx U^{\alpha \beta}
\label{def_rcond_wei}
\end{eqnarray}
Applying the same transformations $ i \mapsto j $ and $ \alpha \mapsto \beta $ we get: 
\begin{eqnarray}
U^{\alpha \beta} \langle w_{ij}^{\beta} \rangle & \approx & \langle \min \{ w_{ij}^{\alpha}, w_{ij}^{\beta} \} \rangle = \nonumber \\
                                                & =    & \langle \min \{ w_{ij}^{\beta}, w_{ij}^{\alpha} \} \rangle \approx U^{\beta \alpha} \langle w_{ij}^{\alpha} \rangle
\label{star_wei}
\end{eqnarray}
Summing (\ref{star_wei}) over $ i $ and $ j $, we have: 
\begin{eqnarray}
U^{\alpha \beta} W^{\beta} = U^{\beta \alpha} W^{\alpha}
\label{global_W}
\end{eqnarray}
Similarly, inverting (\ref{def_rcond_wei}) we obtain:
\begin{eqnarray}
U^{\alpha \beta} \langle w_{ij}^{\beta} \rangle \approx \langle \min \{ w_{ij}^{\alpha}, w_{ij}^{\beta} \} \rangle
\label{def_rcond_wei_inv}
\end{eqnarray}
and summing the previous expression, as in the binary case:
\begin{eqnarray}
U^{\alpha \beta} & = & \frac{\sum_i\sum_{j < i} \langle \min \{ w_{ij}^{\alpha}, w_{ij}^{\beta} \} \rangle }{\sum_i\sum_{j < i} \langle w_{ij}^{\beta} \rangle } = \nonumber \\
                 & = & \frac{\sum_i\sum_{j < i} \min \{ w_{ij}^{\alpha}, w_{ij}^{\beta} \} }{W^{\beta} }
\end{eqnarray}
Therefore we get:
\begin{eqnarray}
\frac{2}{m_{w}^{\alpha \beta}} = \frac{2 \left( W^{\alpha} + W^{\beta} \right) }{2 \sum_i\sum_{j < i} \min \{ w_{ij}^{\alpha}, w_{ij}^{\beta} \} } = \frac{1}{U^{\alpha \beta}} + \frac{1}{U^{\beta \alpha}}
\end{eqnarray}
where $ m_{w}^{\alpha \beta} $ represents the entry of the weighted multiplexity matrix and $ U^{\alpha \beta} $ is derived from the empirical relationship 
between $ s_i^{\beta} $ and $ s_i^{\alpha \rightrightarrows \beta} $.
In analogy with the binary case, $ m_{w}^{\alpha \beta} $ is therefore the harmonic mean of the conditional probabilities $ U^{\alpha \beta} $ and $ U^{\beta \alpha} $, as previously defined. 
Applying (\ref{global_W}) to the previous expression, we get:
\begin{eqnarray}
\frac{2}{m_{w}^{\alpha \beta}} & = & \frac{1}{U^{\alpha \beta}} \left( 1 + \frac{U^{\alpha \beta}}{U^{\beta \alpha}} \right) = \nonumber \\
                                 & = & \frac{1}{U^{\alpha \beta}} \left( 1 + \frac{W^{\alpha}}{W^{\beta}} \right) = \nonumber \\
                                 & = & \frac{W^{\alpha} + W^{\beta}}{U^{\alpha \beta} W^{\beta}}
\end{eqnarray}
Thus, the value of the angular coefficient in the plots $ s_i^{\alpha \rightrightarrows \beta} $ vs $ s_i^{\beta} $ should be, in the weighted case:
\begin{eqnarray}
U^{\alpha \beta} = \frac{W^{\alpha} + W^{\beta}}{2 W^{\beta}} m_{w}^{\alpha \beta}
\label{rel_U_mul_und_wei}
\end{eqnarray}
in perfect analogy with the unweighted case.
Indeed, in Figure~\ref{fig:strength_mul_und} we show the comparison between the actual fit lines and the expected ones according to (\ref{rel_U_mul_und_wei}): the agreement is clear and robust across different pairs of commodities.

Therefore, in analogy to the unweighted case, here the minimal model suitable to reproduce the observed values of pairwise weighted multiplexity is based on the 
total multiplexed weight 
$ W^{\alpha \rightrightarrows \beta} $ for any ordered pair of layers $ (\alpha,\beta) $, accompanied by the strength sequences measured in any layer. We indeed show that any model that does not take into account some sort of weighted coupling between layers would not be sufficient, as shown by the results provided by the uncorrelated model (yellow dots in Figure~\ref{fig:strength_mul_und}).

\subsection{Directed weighted model}

Also in the weighted case it is possible to extend the analysis to the directed case. Here, the main goal consists in the study of the relation between single-layer metrics and inter-layer weighted quantities, in order to model them exploiting the notions of directed multiplexity and multireciprocity introduced before.

We have to define two distinct strengths, namely the out-strength:
\begin{eqnarray}
s_i^{\alpha,out} = \sum_{j \neq i}^{} w_{ij}^{\alpha}
\label{outstrength_seq_emp_dir}
\end{eqnarray}
and the in-strength:
\begin{eqnarray}
s_i^{\alpha,in} = \sum_{j \neq i}^{} w_{ji}^{\alpha}
\label{instrength_seq_emp_dir}
\end{eqnarray}
Moreover, also the multiplex quantities will split into two separate metrics, i.e. the \emph{multiplexed strength}:
\begin{eqnarray}
s_i^{\alpha \rightrightarrows \beta} = \sum_{j \neq i}^{}  w_{ij}^{\alpha \rightrightarrows \beta} \equiv \min \{ w_{ij}^{\alpha}, w_{ij}^{\beta} \}
\label{strength_seq_mul_emp_dir}
\end{eqnarray}
and the \emph{multireciprocated strength}:
\begin{eqnarray}
s_i^{\alpha \rightleftarrows \beta} = \sum_{j \neq i}^{}  w_{ij}^{\alpha \rightleftarrows \beta} \equiv \min \{ w_{ij}^{\alpha}, w_{ji}^{\beta} \}
\label{strength_seq_rec_emp_dir}
\end{eqnarray}
where $ w_{ij}^{\alpha \rightrightarrows \beta} $ is the multiplexed component of the weights associated to the directed links from $ i $ to $ j $ in layers $ \alpha $ and $ \beta $, and $ w_{ij}^{\alpha \rightleftarrows \beta} $ is the reciprocated component.
$ s_i^{\alpha \rightrightarrows \beta} $ and $ s_i^{\alpha \rightleftarrows \beta} $ are the metrics that will allow us to analyse and model the inter-layer coupling of the weighted World Trade Multiplex.

\begin{figure*}[ht]
\begin{center}
\includegraphics[width=1.0\textwidth]{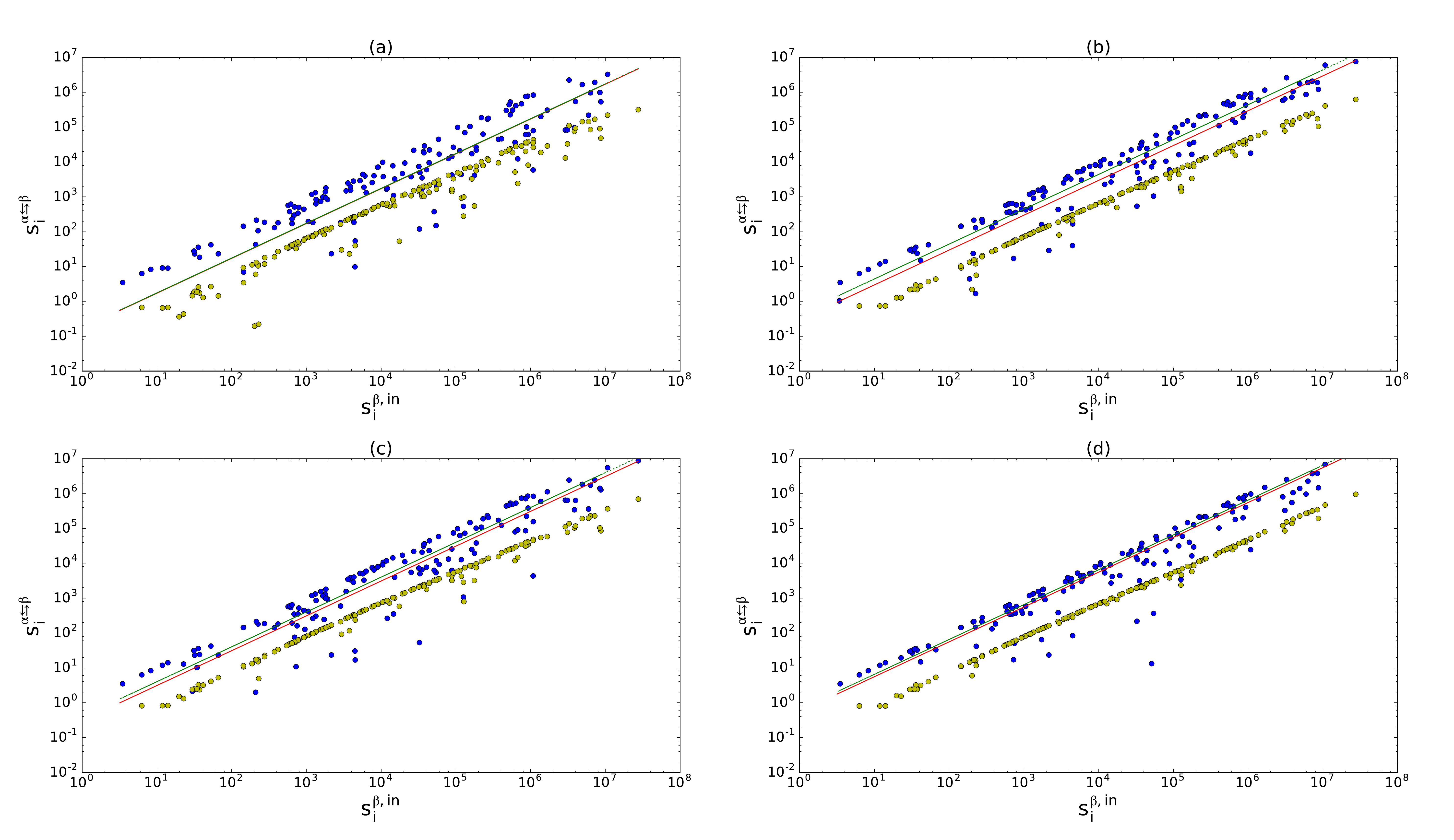}
\end{center}
\caption{In-strength of layer $ \beta $ versus inter-layer multireciprocated strength for 4 different pairs of commodities: inorganic chemicals (a), plastics (b), iron and steel (c), electric machinery (d)
versus trade in cereals. Blue dots: real data; yellow dots: expected multireciprocated strength according to the uncorrelated model; lower green line: expected trend according to (\ref{rel_r_rbin}); upper red line (when discernible): best fit. In all the cases, $ R^2 > 0.95 $, for both the curves. It should be noted that we fit the empirical data with lines of the form $ y = a \cdot x $, and only after we plot the results in log-log scale.}
\label{fig:strength_rec_dir}
\end{figure*}

We empirically observe that the relations between $ s_i^{\alpha,out} $ and $ s_i^{\alpha \rightrightarrows \beta} $ (not shown), and $ s_i^{\alpha,in} $ and $ s_i^{\alpha \rightleftarrows \beta} $ (Figure~\ref{fig:strength_rec_dir}, blue points) are both linearly approximated; hence:
\begin{equation}
s_i^{\alpha \rightrightarrows \beta} \approx U^{\alpha \beta} s_i^{\beta,out}
\label{lin_emp-wei_dir_mul}
\end{equation}
and 
\begin{equation}
s_i^{\alpha \rightleftarrows \beta} \approx V^{\alpha \beta} s_i^{\beta,in}.
\label{lin_emp-wei_dir_rec}
\end{equation}
With the same reasoning developed for the undirected case, it is possible to derive the expected value of the angular coefficient $ U^{\alpha \beta} $ and $ V^{\alpha \beta} $, exploiting the notion of conditional probability; we obtain that the model predicts:
\begin{eqnarray}
U^{\alpha \beta} = \frac{W^{\alpha} + W^{\beta}}{2 W^{\beta}} m_{w}^{\alpha \beta}
\label{rel_U_mul_dir_wei}
\end{eqnarray}
and
\begin{eqnarray}
V^{\alpha \beta} = \frac{W^{\alpha} + W^{\beta}}{2 W^{\beta}} r_{w}^{\alpha \beta}
\label{rel_U_rec_dir_wei}
\end{eqnarray}
where $ m_{w}^{\alpha \beta} $ and $ r_{w}^{\alpha \beta} $ are the corresponding entries of, respectively, the multiplexity and multireciprocity matrices.
The results of the fit of the model to the World Trade Multiplex are shown in Figure~\ref{fig:strength_rec_dir}. The model is able to satisfactorily reproduce the values of multireciprocated strengths starting from the single-layer in-strengths (similar results are obtained for the relation between the out-strength and the multiplexed strength), while an uncorrelated model (i.e., without introducing any sort of dependency between layers) cannot capture the phenomenological observation.

Hence, in the weighted directed case the most inexpensive reconstruction model builds on the knowledge of the in- and out-strength sequence of the different layers plus the $ M \times M $ multiplexity and multireciprocity matrices.


\section{Conclusions\label{sec:conclusions}}

The reconstruction of multiplex properties in multi-layer networks from single-layer information is an important and so far unfaced problem. Indeed, in the multiplex case the limitedness of information about the full topology may affect only some of the layers; hence, any tool allowing us to infer inter-layer node-specific properties from the known information related to some particular layer is theoretically interesting and practically useful. In this article we have provided a possible solution to this issue by means of the new quantities dubbed \emph{multiplexed} and \emph{multireciprocated degrees and strengths}, directly stemming from the previously defined \emph{multiplexity} and \emph{multireciprocity}.
Our reconstruction technique builds on methods that have been shown to be well-grounded in the single-layer case. Indeed, previous studies highlighted that it is possible to correctly reproduce the topological structure of real-world graphs starting from limited information about, for instance, the strengths and the density of the considered system.

In this article we have extended the notion of network reconstruction to the case of multi-layer systems, in particular proving that a trustworthy reconstruction method can be based on the knowledge of (possibly in turn reconstructed) degrees or strengths of the single layers, combined with the compact and usually fixed-over-time multiplexity and multireciprocity matrices. Furthermore, our methodology works for both binary and weighted networks and it is able to take into account also the potential directionality of the links.

We must however stress that this technique is successfully applicable to systems exhibiting two main features. First, the single layers should be reproducible via the Configuration Model (or the Enhanced Configuration Model in the weighted case), such that the entire topology could be reconstructed just from the knowledge of the degrees of the single nodes (respectively, from the strengths). Second, the conditional probabilities of  observing a link in any layer given that a link exists between the same pair of nodes in a different layer should be independent of the considered nodes: in other words, such probabilities (that we called \emph{multiplexity and multireciprocity probabilities}) should be common for all the nodes and dependent only on the pair of layers we are focusing on.
Although these assumptions significanly restrict the range of systems that can be successfully reconstructed through our method, we highlight that one of most crucial economic networks, namely the World Trade Multiplex (incidentally, strongly suffering of the problem of missing data), belongs to this class of multi-layer networks.

Moreover, we have shown that the measures of multiplexed or multireciprocated degrees and strengths can give information about the coupling between layers.
We have indeed explained that, by means of the aforementioned quantities, it is possible to acquire more refined notions of inter-layer coupling; multiplexed and multireciprocated degrees and strengths can therefore be thought of as new measures of multiplex assortativity, expressing the coupling caused by dependencies different than the simple correlation between the degree or strength distributions.

Future steps in the design of reconstruction techniques are needed in order to further generalize the aforementioned methods.
Nevertheless, our findings show that the multiplexity and multireciprocity matrices allow us to reconstruct the joint connection probabilities from the marginal ones, hence bridging the gap between single-layer information and truly multiplex properties.

\end{document}